\documentclass[aps,pre,nofootinbib,twocolumn,superscriptaddress,
               amsmath,amssymb,floatfix]{revtex4}

\usepackage{amsmath} \usepackage{epsfig} \usepackage{bm}

\def\bc{\begin{center}}
\def\ec{\end{center}}
\def\be{\begin{equation}}
\def\ee{\end{equation}}

\def\bear{\begin{eqnarray}}
\def\eear{\end{eqnarray}}

\begin{document}

\title{Lengthscale dependence of dynamic four-point susceptibilities
in glass formers}

\author{David Chandler}

\affiliation{Department of Chemistry, University of California,
Berkeley, CA 94720-1460}

\author{Juan P. Garrahan}

\affiliation{School of Physics and Astronomy, University of
Nottingham, Nottingham, NG7 2RD, UK}

\author{Robert L. Jack}

\affiliation{Department of Chemistry, University of California,
Berkeley, CA 94720-1460}

\author{Lutz Maibaum}

\affiliation{Department of Chemistry, University of California,
Berkeley, CA 94720-1460}

\author{Albert C. Pan}

\affiliation{Department of Chemistry, University of California,
Berkeley, CA 94720-1460}

\begin{abstract}
Dynamical four-point susceptibilities measure the extent of spatial
correlations in the dynamics of glass forming systems. We show how
these susceptibilities depend on the length scales that necessarily
form part of their definition.  The behaviour of these
susceptibilities is estimated by means of an analysis in terms of
renewal processes within the context of dynamic facilitation. The
analytic results are confirmed by numerical simulations of an
atomistic model glass-former, and of two kinetically constrained
models. Hence we argue that the scenario predicted by the dynamic
facilitation approach is generic.
\end{abstract}

\maketitle

\section{Introduction}
\label{sec:intro}

The length scales governing dynamical heterogeneity in glass-forming
liquids
\cite{ReviewsDH,YamamotoO98b,BennemannDBG99,GarrahanC02,BiroliB04} are
often described in terms of the susceptibility associated with
fluctuations in the self intermediate scattering function
\cite{FranzDPG99,GarrahanC02,Ber04,ToninelliWBBB05,cosine-footnote}:
\begin{equation}
\chi_4(k,t) \equiv \frac{1}{N} \sum_{jl} \langle \delta
\hat{F}_j(\bm{k},t) \delta \hat{F}_l(-\bm{k},t) \rangle .
\label{equ:def_chi4}
\end{equation}
Here the indices $j$ and $l$ run over the $N$ particles in the system,
the position of the $j$th particle at time $t$ is $\hat{\bm{r}}_j(t)$;
\begin{equation}
\delta \hat{F}_j(\bm{k},t)\equiv e^{i\bm{k}\cdot[\hat{\bm{r}}_j(t)-
\hat{\bm{r}}_j(0)]}- \langle
e^{i\bm{k}\cdot[\hat{\bm{r}}_j(t)-\hat{\bm{r}}_j(0)]} \rangle ,
\end{equation}
and $k=|\bm{k}|$.  Under supercooled conditions, this four-point
correlation function typically grows in time towards a peak, before
decreasing at large times. This non-monotonic behaviour is a
consequence of the transient nature of dynamic heterogeneity.

It was suggested by Toninelli~\emph{et~al.}  \cite{ToninelliWBBB05}
that the time dependence of $\chi_4(k,t)$ can be used to distinguish
between different theoretical scenarios for the glass transition.  The
dependence of $\chi_4(k,t)$ on the wave vector $k$ was considered for
a glass-forming system in \cite{Lacevic}, and for a sheared granular
material in \cite{DauchotMB05}.  In both cases, significant dependence
on wave vector was found.  In this article, we investigate this wave
vector dependence, in particular the way that $\chi_4(k,t)$ grows
towards its peak.  We present data for an atomistic system, and for
two kinetically constrained models \cite{RitortS03}. We find
non-trivial wave vector dependence in all three cases.  We explain
this generic behaviour analytically using a treatment that we used
earlier to describe dynamic decoupling in glass formers
\cite{JungGC04,JungGC05,BerthierCG05}.

Our analysis shows that the non-trivial behaviour of four-point
correlators comes from two sources. The first contribution arises
because particles that have not moved are clustered in space; the
second comes from correlations between particle displacements.  The
relative sizes of these contributions depend on the wave vector
$k$. The first dominates when $k$ is large, and the second dominates
when $k$ is small.  The crossover between these two regimes
corresponds to the crossover between non-Fickian and Fickian regimes
observed in two-point functions \cite{BerthierCG05}.

\begin{figure}
\epsfig{width=0.95\columnwidth,file=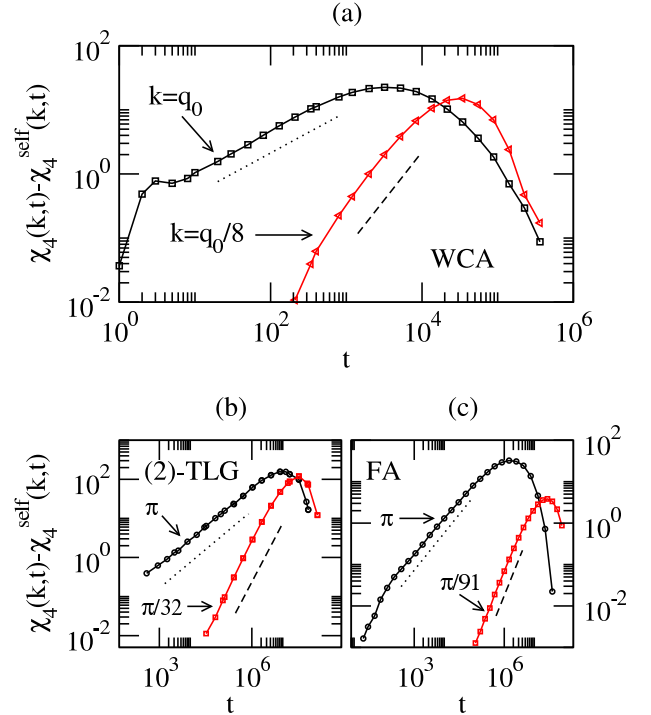}
\caption{(Color online) The distinct part of $\chi_4(k,t)$ as a function of $t$ for
two wave vectors in three model systems: the supercooled three
dimensional WCA mixture (a); a kinetically constrained triangular
lattice gas (b); and probes in the one dimensional
Fredrickson-Andersen model (c).  The symbol
$\chi_4^\mathrm{self}(k,t)$ denotes $\langle |\delta \hat{F}_i(k,t)|^2
\rangle$.  Full details are presented in the relevant sections below.
Dashed and dotted lines show that in all cases the exponents of power
law fits increase with decreasing wave vector. The largest wave
vectors considered are the principal wave vectors for each system (for
the WCA case this is the peak location in the equilibrium structure
factor, $q_0$).  }
\label{fig:chi4}
\end{figure}

Figure~\ref{fig:chi4} illustrates the behaviour that we consider for
three different model systems.  These systems, described in detail in
sections~\ref{sec:wca} and \ref{sec:kcm}, represent three levels of
coarse-graining in the glassy system. The most detailed is a fluid
mixture of classical particles in continuous three dimensional space,
interacting with Weeks-Chandler-Andersen (WCA) potentials
\cite{WCA71,Maibaum05}.  The second is the so-called (2)-TLG, due to
J\"{a}ckle and Kr\"{o}nig \cite{JackleK94}. It is a kinetically
constrained lattice gas in which the dynamics are highly co-operative
and relaxation times increase very quickly with increasing density.
The final system is the one dimensional one spin facilitated
Fredrickson-Andersen (FA) model \cite{FredricksonA84} to which probe
particles have been added, following Ref.\ \cite{JungGC04}.  The probe
particles do not interact with one another, but they propagate through
an environment that is dynamically heterogeneous.  In this model, all
of the atomistic interactions have been removed, leaving only an
idealised dynamically heterogeneous system.

The similarity of $\chi_4(k,t)$ between these three different model
systems is striking.  The four-point susceptibility has the usual
single peak, and the increasing part of each curve can be fitted by a
power law. In all three cases, the exponent with which $\chi_4(k,t)$
increases depends strongly on $k$. Further, all cases exhibit a shift
of the peak of $\chi_4(k,t)$ to later times as $k$ decreases.

\section{Theory of four-point functions in heterogeneous systems}
\label{sec:analysis}

\subsection{The dynamical facilitation approach}

Our starting point is to follow \cite{BerthierCG05} and write
\begin{eqnarray}
\hat{F}_j(\bm{k},t) & \equiv & \exp[i\bm{k}\cdot \Delta \hat{\bm{r}}_j(t)]
\nonumber\\
& = & \hat{p}_j(t) + \left[ 1 - \hat{p}_j(t) \right]
\exp{\left[ i \bm{k} \cdot \Delta \hat{\bm{r}}_j(t) \right]} ,
\label{Fsump}
\end{eqnarray}
where $\Delta \hat{\bm{r}}_j(t)\equiv
\hat{\bm{r}}_j(t)-\hat{\bm{r}}_j(0)$ and $\hat{p}_j(t)$ is the local
persistence operator. That is, $\hat{p}_i(t)$ takes the value of unity
if particle $j$ has not moved a distance greater than some microscopic
cutoff $a_0$, and it is zero otherwise.  We use hats throughout this
article to denote fluctuating quantities (operators).

The usefulness of (\ref{Fsump}) lies in the fact that the two terms
separate mobile and immobile particles, explicitly accounting for the
dynamical heterogeneity in the system. The expectation of each term
can be simply evaluated in an appropriate Gaussian, or homogeneous,
approximation, leading to:
\begin{eqnarray}
F_{\rm s}(k,t) & \equiv & \langle \hat{F}_j(\bm{k},t) \rangle 
\nonumber \\ & \approx & P(t) + [1
- P(t)] \exp{(-k^2 D t)} ,
\label{Fs}
\end{eqnarray}
where $P(t)$ is the average persistence function $P(t) \equiv \langle
\hat{p}_j(t) \rangle$ and $D$ is the self-diffusion constant. The
approximate equality is valid \cite{BerthierCG05} in the deeply
supercooled regime of large decoupling between $\alpha$-relaxation
time and diffusion rate \cite{SEpapers}. According to this
approximation, particle motion is a random walk with randomly
distributed waiting times \cite{MW65}.

Moving from two point to four point functions, we consider the
correlator $G_{jl}(k,t)=\langle \delta\hat{F}_j(\bm{k},t)
\delta\hat{F}_l(-\bm{k},t) \rangle$ where $\delta \hat{F}_j =
\hat{F}_j - F_{\rm s}$ was defined above.  The diagonal (self) part
is:
\begin{equation}
G_{jj}(k,t) = 1 - F_{\rm s}(k,t)^2
.
\label{chis}
\end{equation}
It grows monotonically with time and is always smaller than unity.
Non-trivial spatial correlations of the dynamics appear in the
off-diagonal terms $j \neq l$.  Using Eqs.\ (\ref{Fsump}) and
(\ref{Fs}), these correlations are
\begin{eqnarray}
\label{dFdF}
\lefteqn{ G_{jl}(k,t) \approx \left( 1 - e^{-k^2 D t} \right)^2
\langle \delta \hat{p}_j(t) \delta \hat{p}_l(t) \rangle } \qquad\qquad
\nonumber \\ && + \left[ 1 - P(t) \right]^2 \langle \delta
e^{i\bm{k}\cdot \Delta \hat{\bm{r}}_j(t)} \, \delta e^{ -i\bm{k} \cdot
\Delta \hat{\bm{r}}_l(t)} \rangle,
\end{eqnarray}
where $\delta \hat{p}_j = \hat{p}_j - P$ and $\delta e^{ i \bm{k}
\cdot \Delta \hat{\bm{r}}_j(t)} = e^{ i\bm{k} \cdot \Delta
\hat{\bm{r}}_j(t)} - e^{-k^2 D t}$. We have dropped terms that are
cubic and quartic in the fluctuations since we expect their
contributions to be small.

\begin{figure*}
\epsfig{file=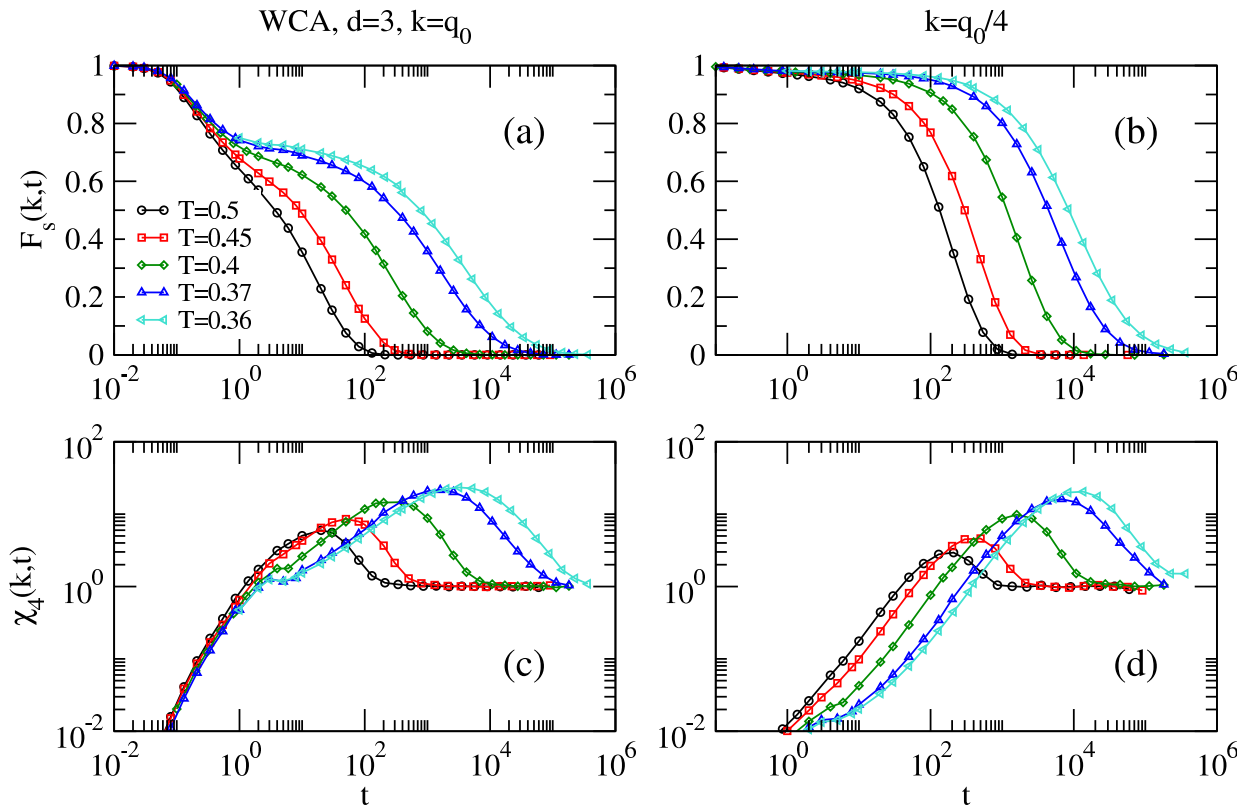,width=1.9\columnwidth}
\caption{(Color online) (a,b) Self-intermediate scattering functions
  $F_s(k,t)$ in the WCA mixture for wave vectors $k=q_0$ and $k=q_0/4$
  ($q_0$ is the wave vector of the first peak in the structure factor).
  (c,d) Corresponding four-point susceptibilities $\chi_4(k,t)$.}
\label{fig:atom_temp}
\end{figure*}

To arrive at the four-point susceptibility, we must sum over $j$ and
$l$.  The first term of (\ref{dFdF}) is the susceptibility of the
persistence, and we have
\begin{equation}
\frac{1}{N} \sum_{jl} \langle \delta \hat{p}_j(t) \delta \hat{p}_l(t)
\rangle \approx N_{\rm p}(t) P(t) \left[ 1 - P(t) \right] .
\label{chip}
\end{equation}
Here we have assumed that excitations propagate through the system
leading to movement of particles \cite{JungGC04}, and that the average
number of particles visited by a single excitation in time $t$ is
$N_{\rm p}(t)$.  This number is related to the dynamic correlation
length of relaxed clusters of particles through an exponent
representing their (possibly fractal) dimensionality.  $N_{\rm p}(t)$
is also is related to the average persistence: a molecule will
typically persist after time $t$ if there were no excitations
initially in a surrounding region of mass $N_{\rm p}(t)$; the
probability for this to happen is $P(t) \approx e^{-c N_{\rm p}(t)}$,
where $c \ll 1$ is the average concentration of excitations.  The
calculation of the average of the product of the persistence is
analogous: for $p_j p_l$ to be nonzero after time $t$ a mass $N_{\rm
p}(t)$ has to be free of excitations initially around $j$ and $l$.
The fact that these two volumes may overlap gives rise to spatial
correlations between $p_j$ and $p_l$.  Following similar arguments to
those used in Ref.\ \cite{ToninelliWBBB05} one arrives to equation
(\ref{chip}).

While the contribution to $\chi_4(k,t)$ from equation (\ref{chip})
measures whether nearby particles relax in a correlated way, the
second term of (\ref{dFdF}) measures correlations in the displacements
of particles that have relaxed.  We define the correlation function
for particle displacements in the following way:
\begin{equation}
g(r,t) \equiv t^{-1} \langle \Delta\hat{\bm{r}}_j(t) \cdot
\Delta\hat{\bm{r}}_l(t) \rangle_{r_{jl}=r}.
\label{gdef}
\end{equation}
where the average is conditioned on the initial separation of the
particles $r_{jl}\equiv|\bm{r}_j(0)-\bm{r}_l(0)|$.  This function
$g(r,t)$ measures the correlations between displacements of nearby
particles.  For simple Brownian motion, the separation of a pair of
particles diffuses four times as fast as their centre of mass. If
$g(r,t)$ is positive, then the separation diffuses more slowly than
this reference value; if it is negative then the separation diffuses
faster.  If two particles move through the system facilitated by the
same excitation then their separation remains relatively small, while
their centre of mass moves a long way.  We show the resulting positive
$g(r,t)$ for a simple kinetically constrained model in Fig 6 (below).

Within the Gaussian approximation, we arrive at
\[
\langle \delta e^{ i \bm{k} \cdot \Delta \bm{r}_j(t)} \delta e^{- i 
\bm{k}
\cdot \Delta\hat{\bm{r}}_l(t)} \rangle_{r_{jl}=r}
\approx e^{-2 k^2 D t} \left( e^{ 2 t
k^2 g(r,t) } - 1 \right) .
\]
The Gaussian approximation is justified because particles have made
many diffusive steps.  The sum over particles at a given time $t$ is
dominated by pairs of particles whose initial separation coincides
with the maximum of $g(r,t)$.  This leads to a contribution to
$\chi_4(k,t)$ of
\begin{eqnarray}
\frac{1}{N} \sum_{jl}
\langle \delta e^{i\bm{k}\cdot \Delta \hat{\bm{r}}_j(t)} \, \delta e^{ -i\bm{k}
\cdot \Delta \hat{\bm{r}}_l(t)} \rangle \qquad\qquad\nonumber\\
\approx N_g(t) \, e^{-2 k^2 D t} \left( e^{ 2 t k^2 \bar{g}(t) } - 1
\right) ,
\label{dFtdFt}
\end{eqnarray}
where $\bar{g}(t)=\max_x[g(x,t)]$ obeys $\bar{g}(t)<D$ and decreases
at large times.  We have assumed that the sum over $j$ is dominated by
particles for which $g(r,t) \simeq \bar{g}(t)$, and we denote the
number of these particles by $N_g(t)$.  The summand is exponential in
$t \, g(r,t)$, so this is a good approximation when that function is
large. In the excitation picture we only expect strong displacement
correlations when two particles are facilitated by the same
excitation, so we expect $N_g(t)<N_p(t)$ in general.

Putting the results of Eqs.\ (\ref{chis})-(\ref{chip}) and
(\ref{dFtdFt}) together we arrive at the four-point susceptibility,
\begin{eqnarray}
\label{chik}
\chi_4(k,t) &\approx& N_\mathrm{p}(t) P(t) \left[ 1 - P(t) \right]
\left( 1 - e^{-k^2 D t} \right)^2 \nonumber \\ && + N_g(t) \, \left[ 1
- P(t) \right]^2 e^{-2 k^2 D t} \left( e^{ 2 t k^2 \bar{g}(t) } - 1
\right) \nonumber \\ && + 1 - F_{\rm s}(k,t)^2 .
\end{eqnarray}

Thus, correlations between particles come from persistence
correlations [first term in (\ref{chik})] and from displacement
correlations [second term in (\ref{chik})].  Both contributions are
non-monotonic in time. The persistence contribution peaks at a time
$t_{\rm peak}$ that scales as $t_{\rm peak} \sim \tau_\alpha$, where
$\tau_\alpha$ is the structural relaxation time.  This term is
relevant if $k$ is large.  Conversely, the second term in (\ref{chik})
dominates at small $k$, when $Dk^2\tau_\alpha\ll1$. The peak of this
term occurs at $t_{\rm peak} \sim (D k^2)^{-1}$, and so increases with
decreasing $k$, while its peak height decreases as $\bar{g}(t_{\rm
peak})$.

Equation~(\ref{chik}) is consistent with the data of Fig.\
\ref{fig:chi4}. As well as the overall form of $\chi_4$, it contains
two main predictions.  Firstly, for a given wave vector, $\chi_4(k,t)$
peaks at a $k$-dependent time $t_{\rm peak}$. The scaling of this time
is the same as that of the time scale of two-point correlators
\cite{BerthierCG05}.  It increases with decreasing $k$, consistent
with simulations.  A corollary is that $\chi_4(k,t)$ may display a
non-trivial structure even for wavelengths at which the corresponding
one-particle motion is Fickian.

\begin{figure*}
\epsfig{file=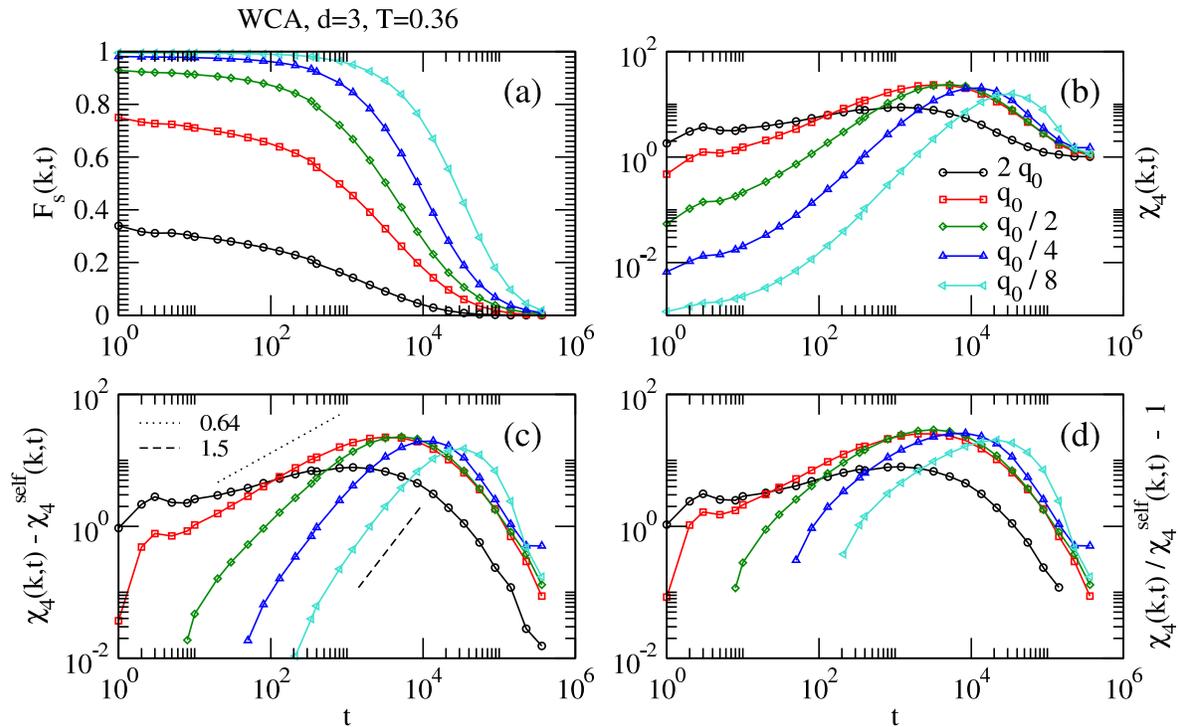,width=1.8\columnwidth}
\caption{(Color online)  (a) Self-intermediate scattering functions $F_s(k,t)$ in the
WCA mixture at $T=0.36$, for wave vectors $k=2 q_0, q_0, q_0/2, q_0/4,
q_0/8$.  (b) Corresponding four-point susceptibilities $\chi_4(k,t)$.
(c) Distinct part $\chi_4(k,t) - \chi_4^{\rm self}(k,t)$.  (d)
Distinct part normalized by the self term, $\chi_4(k,t)/\chi_4^{\rm
self}(k,t)-1$.  } \label{fig:atom_k}
\end{figure*}

\begin{figure}
\epsfig{width=0.95\columnwidth,file=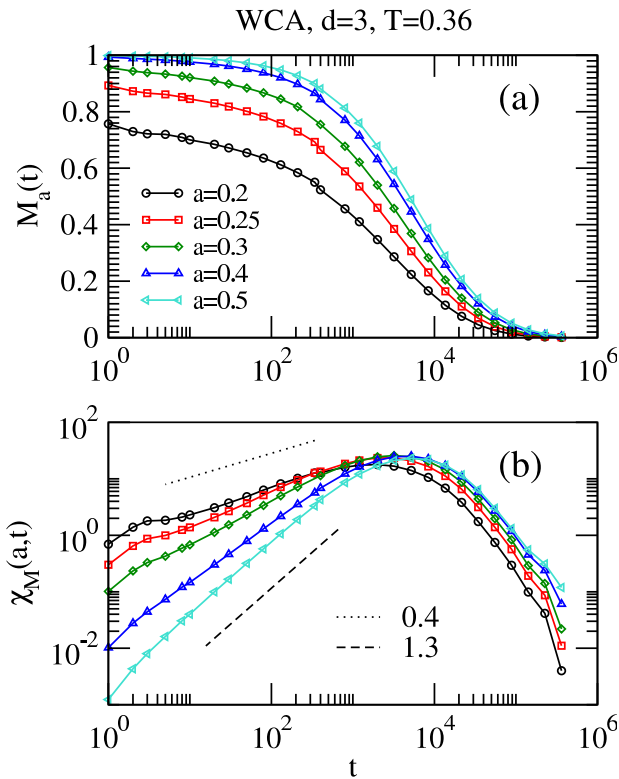}
\caption{(Color online)  (a) Two point overlap correlators $M_a(t)$ in the WCA
atomistic model at temperature $T=0.36$.  (b) Corresponding four point
susceptibilities $\chi_M(a,t)$.  }
\label{fig:chiM}
\end{figure}

Secondly, the increase of $\chi_4(k,t)$ depends on the wave vector $k$:
from above, we have for large $k$
\begin{equation}
\chi_4(k,t) \sim
 N_\mathrm{p}(t) \left[1-P(t)\right], \quad Dk^2t \gg 1 
,
\label{equ:chi4_scaling_largek}
\end{equation}
and for small $k$
\begin{equation}
\chi_4(k,t)\sim N_g(t) \left[1-P(t)\right]^2 \overline{g}(t) k^2 t, 
\quad Dk^2t \ll 1.
\label{equ:chi4_scaling_smallk}
\end{equation}
The two relations come from persistence (large $k$) and displacement
correlations (small $k$).  These two scaling predictions are different
in general: we argue that this accounts for the variation with $k$ at
early times that was demonstrated in figure~\ref{fig:chi4}.  Within
the simplest picture of facilitation, $N_p(t)$ is the fundamental
object: it measures the number of particles whose relaxation is
facilitated by a single mobility excitation. At this minimal level of
theory, and for small times ($t<<\tau_\alpha$), we expect the number
of particles that have relaxed to be $N_c N_p(t)$, where $N_c$ is the
number of mobility excitations; it follows that $[1-P(t)] \sim
N_p(t)$.  Of the $N_p$ particles facilitated by a single excitation, we
expect a finite fraction to have strong displacement correlations:
this finite fraction is $N_g/N_p$. If this assumption holds then
$N_g(t)$ and $N_p(t)$ will have the same scaling with time and
temperature.

Following Toninelli~\emph{et~al.}~\cite{ToninelliWBBB05}, we define an
exponent $\mu$ by $\chi_4(t) \sim t^\mu$. The observation of
Ref.~\cite{ToninelliWBBB05} is that $\mu\simeq b$, where $b$ is the
von-Schweidler $\beta$-relaxation exponent of mode-coupling theory
(MCT) \cite{Got99}.  Eq.\ (\ref{chik}) shows clearly that $\mu$
depends on wave vector, $\mu=\mu(k)$; however, MCT predicts a
dynamical critical point at which $b$ is independent of $k$
\cite{Got99} (see nevertheless the discussion in \cite{FGM98}).  In
supercooled liquids, no such phase transition is observed, and the
system is always in the ergodic phase. Hence, all correlation lengths
are finite, and the scaling relation $\mu\simeq b$ breaks down for
wave vectors smaller than the inverse correlation length. This effect
can be investigated by direct numerical solutions of the MCT equations
\cite{Foffi,FlennerS05}, but agreement with simulation is still poor
for small wave vectors. This is consistent with the hypothesis that
processes neglected by the mode coupling approximation are important
for the structural relaxation of supercooled liquids, leading to
avoidance of the dynamical transition; to the decoupling of diffusion
and viscosity \cite{FlennerS05}; and to non-trivial wave vector
dependence of two and four-point correlations on large length scales.

\subsection{Related four-point functions}

The function
\[
G_4(\bm{k},\bm{q},t) = \frac{1}{N}
\sum_{jl}\langle \delta \hat{F}_j(\bm{k},t) \delta \hat{F}_l(-\bm{k},t) 
\mathrm{e}^{i\bm{q}\cdot [\hat{\bm{r}}_j(0)-\hat{\bm{r}}_l(0)]} \rangle
\]
generalizes $\chi_4(k,t)$. Clearly $\chi_4(k,t)=G_4(\bm{k},0,t)$.
However, $G_4(\bm{k},\bm{q}=0,t)$ is ensemble dependent (see e.g.
\cite{BerthierBBCMLLP05}), whereas $\lim_{q\to0} G_4(\bm{k},\bm{q},t)$
is not. For the WCA mixture and the (2)-TLG we have found that the
ensemble dependence affects the absolute value of $\chi_4(k,t)$ but
not their functional trends.  The differences in $\chi_4(k,t)$ between
ensembles can be calculated in terms of thermodynamic properties and
derivatives of $F_\mathrm{s}(k,t)$ \cite{LebowitzPV67}.
Berthier~\emph{et al.}~\cite{BerthierBBCMLLP05} suggest that these
differences provide reliable estimates of $\lim_{q\to0}
G_4(\bm{k},\bm{q},t)$ itself.  For the WCA mixture and (2)-TLG, in the
regimes accessible to our simulations, we find that these differences
also have an important $k$ dependence, and that they only account for
a fraction of the total value of $\lim_{q\to0} G_4(\bm{k},\bm{q},t)$.
This is discussed in detail in the Appendix.

An alternative (ensemble dependent) four point function is
\begin{equation}
\chi_M(a,t) \equiv \frac{1}{N} \sum_{jl} 
\langle \delta\hat{M}_j(a,t) \delta\hat{M}_l(a,t) 
\rangle
\end{equation}
where $\delta\hat{M}_i=\hat{M}_i-\langle \hat{M_i}\rangle$ in which
$\hat{M}_i(a,t)$ is a (binary) operator that equals unity if
$|\hat{r}_i(t)-\hat{r}_i(0)|<a$ and zero otherwise.

The operator $\hat{M}(a,t)$ was used, for example, in
\cite{Lacevic}.  Its ensemble average, $M_a(t) \equiv \langle
\hat{M}_j(a,t) \rangle$, is the fraction of particles that have not
moved beyond a distance $a$ in time $t$, and is related to the
self-intermediate scattering function by $M_a(t)=(2\pi)^{-d}
\int_{|\bm{r}|<a}\!  \mathrm{d}^d \bm{r} \int\!\mathrm{d}^d\bm{k}\,
e^{-i\bm{k}\cdot\bm{r}} F_\mathrm{s}(k,t)$.

Our analysis of $\chi_4(k,t)$ 
generalizes immediately to $\chi_M$: we write
\begin{equation}
\hat{M}_j(a,t) = \hat{p}_j(t) + \left[ 1 - \hat{p}_j(t) \right]
\Theta(a-|\Delta\hat{\bm{r}}_j(t)|)  ,
\label{Msump}
\end{equation}
where $\Theta(x)$ is the Heaviside function.  The difference between
the operators $\hat{M}_j(a,t)$ and $\hat{F}_j(k,t)$ is that the phase
of $\hat{F}_j(k,t)$ records the direction of motion of the $j$th
particle. For small $a$, correlations between the directions in which
mobile particles have moved do not contribute to $\chi_M(a,t)$ and we
expect it to be dominated by the persistence correlations:
\begin{eqnarray}
\label{chim}
\chi_M(a,t) &\approx& f(Dt/a^2)
 N_\mathrm{p}(t) P(t) \left[ 1 - P(t) \right] 
\end{eqnarray}
where $f(x)= (4\pi Dt)^{-d} [\int_{|\bm{r}|<a}\!\mathrm{d}^d\bm{r}\,
e^{-|\bm{r}|^2/(Dt)}]^2$.  The structure is the same as the first term
of (\ref{chik}) with $f(Dt/a^2)$ playing the part of
$(1-e^{-Dk^2t})^2$. We note also that $\chi_M(a,t)$ approaches the
contribution of (\ref{chip}) as $a$ gets small, as expected.  We
compare this prediction with atomistic simulations in
section~\ref{sec:wca}, see below.

\section{Atomistic simulations}
\label{sec:wca}

For a continuous atomistic representation of a supercooled liquid, we
carried out extensive molecular dynamics simulations of a symmetric
WCA mixture \cite{Maibaum05}.  It is a mixture of two particle species
A and B in three spatial dimensions.  The potential energy is the sum
of the pairwise interactions between two particles of species $\alpha$
and $\beta$, $V_{\alpha \beta}(r) = 4 \varepsilon \left[ \left(
\sigma_{\alpha \beta} / r \right)^{12} - \left( \sigma_{\alpha \beta}
/ r \right)^6 + 1/4 \right]$ if $r \leq 2^{1/6} \sigma_{\alpha
\beta}$, and $V_{\alpha \beta}(r) = 0$ otherwise.  This is the
reference potential of the WCA theory \cite{WCA71} consisting of the
repulsive part of the Lennard-Jones interaction.  Following
\cite{Wahnstrom91,Lacevic} we choose: $\sigma_{AA} = 1$,
$\sigma_{BB} = 5/6$, $\sigma_{AB}=(\sigma_{AA}+\sigma_{BB})/2$,
$m_B=m_A/2=1$, and $\varepsilon=1$.  Lengths, times and temperatures
are given in units of $\sigma_{AA}$, $\sqrt{m_B
\sigma^2_{AA}/\varepsilon}$, and $\varepsilon/k_B$, respectively; we
use $q_0$ to denote the wave vector of the first peak in the structure
factor.  In our simulations the total number of particles was
$N=8000$, with $N_A=N_B=4000$.  The use of the WCA reference potential
makes this system computationally more efficient to simulate than the
original Lennard-Jones one of Ref.\ \cite{Wahnstrom91}. A detailed
study of the dynamics in the supercooled regime of this WCA mixture
will be presented elsewhere \cite{WCA06}.  The molecular dynamics
simulations of this section conserve energy, and we use a range of
energies at each temperature. This allows estimates of both canonical
and microcanonical susceptibilities. Details are given in the
Appendix; the data of Figs. \ref{fig:atom_temp}-\ref{fig:chiM} is for
the canonical susceptibility.

Figure~\ref{fig:atom_temp} shows the self-intermediate scattering
function $F_\mathrm{s}(k,t)$ at various temperatures $T=0.5$ to
$0.36$, for the wave vector of the first peak of the structure factor
$k=q_0$, and for a smaller wavevector $k=q_0/4$.  At the lowest
temperature shown, $T=0.36$, the system is clearly in the supercooled
regime, dynamics is heterogeneous, and the self-diffusion constant
exceeds the value expected by the Stokes-Einstein relation by over an
order of magnitude \cite{WCA06}. The four-point susceptibilities have
the expected behaviour, becoming larger and peaking later the lower
the temperature, as shown in the lower panels of
figure~\ref{fig:atom_temp}.

Figure~\ref{fig:atom_k} concentrates on the lowest temperature we
simulated, $T=0.36$.  It shows the two point function
$F_\mathrm{s}(k,t)$ for various wave vectors $k$.  The growth of the
distinct part of $\chi_4$ towards its peak can be fitted by a power of
$t$. The fitted exponent is smaller than $1$ for wave vectors near
$q_0$, but closer to $1.5$ for small $k$.  The former result coincides
with what was observed in similar systems \cite{ToninelliWBBB05}, but
the latter was not anticipated before.  Interestingly, a similar $k$
dependence has been observed experimentally in a sheared granular
material \cite{DauchotMB05}.

In section~\ref{sec:analysis}, we assumed that $\chi_M(a,t)$ is
dominated by persistence correlations for small $a$, where
displacement correlations are unimportant. At small times, the effect
of increasing $a$ is to reduce the susceptibility.  This is analogous
to the effect of decreasing $k$ in $\chi_4(k,t)$.  On the other hand,
the $k$ dependence of the peak in $\chi_4(k,t)$ arises from the
displacement correlations and will be absent (or at least much weaker)
in $\chi_M(a,t)$.  Both these predictions are consistent with the data
of Fig.~\ref{fig:chiM}, calculated in the atomistic system of
section~\ref{sec:wca}. However, we note that the relative range of $a$
over which we have measured $\chi_M(a,t)$ is smaller than the range of
$k$ used for $\chi_4(k,t)$. It may be that particle displacement
correlations will be important at larger $a$.  This effect can be
estimated within the framework of section~\ref{sec:analysis}.  It
appears as an extra term in (\ref{chim}).

\section{Kinetically constrained models}
\label{sec:kcm}

\subsection{Kinetically constrained lattice gas}
\label{subsec:tlg}

\begin{figure}
\epsfig{file=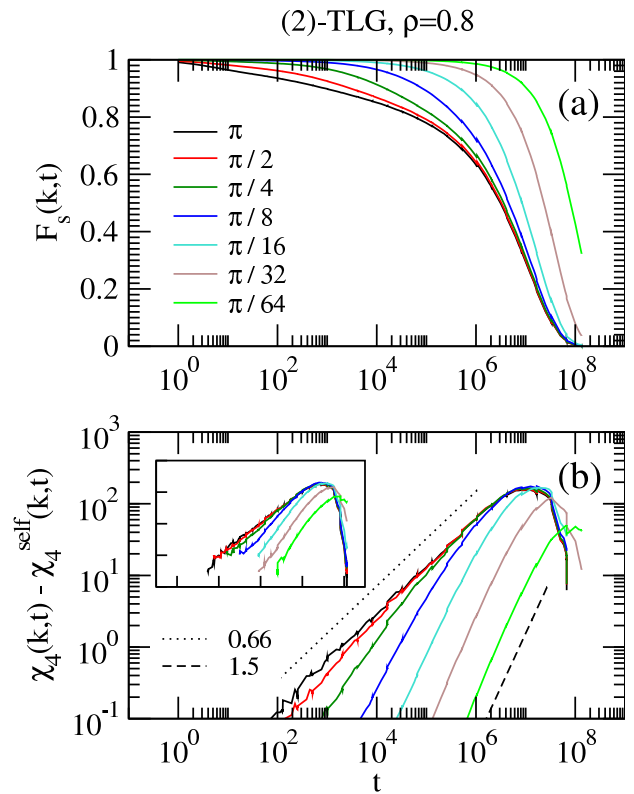,width=0.95\columnwidth}\par
\caption{(Color online)  (a) Self-intermediate scattering function $F_s(k,t)$ of
particles in the (2)-TLG model, at density $\rho=0.8$, for wave
vectors $k=\pi,\pi/2,\pi/8,\pi/16,\pi/32,\pi/64$.  (b)
Susceptibilities $\chi_4(k,t)$ for the self-correlators of the top
panel. For large $k$, $\chi_4$ grows with an exponent close to $0.7$,
for small $k$ with an exponent close to $1.5$.  The inset shows the
normalized function $\chi_4 / \chi_4^{\rm self} -1$ in the same scale
as the main panel.  }
\label{fig:tlg}
\end{figure}

The kinetically constrained lattice gases of J\"{a}ckle and Kr\"{o}nig
\cite{JackleK94} provide simple caricatures of supercooled liquids. We
consider the (2)-TLG in which hard core particles move on a triangular
lattice: movement between sites and $i$ and $j$ is allowed only if
both the mutual neighbours of sites $i$ and $j$ are empty.  Relaxation
in the model involves strongly co-operative motion and the relaxation
time increases very rapidly with increasing density.

We show correlation functions in Fig.~\ref{fig:tlg}.  We plot the
self-intermediate scattering functions for particles at a high density
$\rho=0.8$; we use wave vectors from $k=\pi$ to $k=\pi/64$ (for
details see Ref.\ \cite{PGC05}).  We note that the exponents governing
the growth of $\chi_4(k,t)$ in the (2)-TLG are rather similar to those
in the atomistic system. These four-point functions are measured at
constant density.

\subsection{Fredrickson-Andersen (FA) model}
\label{subsec:fa}

\begin{figure}
\epsfig{file=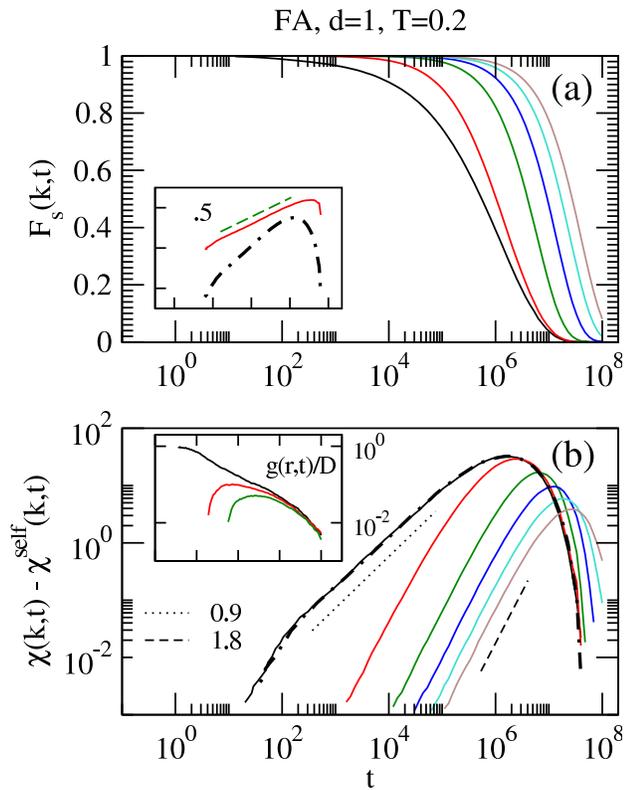,width=0.95\columnwidth}\par
\caption{(Color online) (a) Self-intermediate scattering function
$F_s(k,t)$ of probes embedded in an FA model in $d=1$, at temperature
$T=0.5$, for wave vectors $k=\pi,\pi/11,\pi/31,\pi/51,\pi/71,\pi/91$.  The
inset shows the susceptibility of the persistence given in Eq.\
(\ref{chip}) (dot-dashed) and the quantity $N_\mathrm{p}(t)$ (full),
which grows approximately as $t^{1/2}$ (dashed).  (b) Susceptibilities
$\chi_4(k,t)$ for the self-correlators of the top panel.  We show the
non-trivial part, with the self-term $\chi_4^{\rm self}(k,t)$ removed.
$\chi_4$ for $k=\pi$ follows the contribution of Eq.\ (\ref{chip})
(dot-dashed).  For large $k$, $\chi_4$ grows with an exponent close to
$1$, for small $k$ with an exponent close to $1.5$.  The inset shows
the function $g(r,t)$, see Eq.\ (\ref{gdef}), for inter particle
distances $r=0,1,2$, in the same temporal scale as the main panel.  }
\label{fig:fa}
\end{figure}

The one-spin facilitated Fredrickson-Andersen (FA) model represents
the extreme of coarse-graining in which one conjectures that the only
variables relevant for heterogeneity are binary labels for regions in
which mobility is present.  It is the simplest kinetically constrained
model displaying both dynamic heterogeneity \cite{GarrahanC02} and
decoupling \cite{JungGC04}.  The system is a one dimensional chain in
which mobile regions are represented by up spins and immobile regions
by down spins. The energy of the system is simply the number of up
spins: spins flip with Metropolis rates if and only if at least one
neighbouring spin is up.

In order to consider particle motion in such a system, we couple probe
particles to the spins as in \cite{JungGC04}. Probes can hop only
between adjacent mobile sites and such moves are attempted with rate
unity. Probes do not interact with one another (more than one probe
may occupy a single site). The stationary distribution of probes is
uniform and uncorrelated with the spin variables.

Figure~\ref{fig:fa} (a) shows the self-intermediate scattering
functions of probe molecules in the FA model at a low temperature
$T=0.2$ for various wave vectors (the ensemble is again that in which
the probe density is constant).  For large wave vector the
self-correlator is stretched and tracks the persistence function,
while for small enough wave vectors self-correlators become
exponential, according to Eq.\ (\ref{Fs}) (see Ref.\
\cite{BerthierCG05} for details).  The inset shows the time dependence
of $N_\mathrm{p}(t)$, as extracted from the susceptibility of the
persistence, see Eq.\ (\ref{chip}).  Figure~\ref{fig:fa} (b)
shows the corresponding susceptibilities $\chi_4(k,t)$.  We note that
the exponent governing the increase of $\chi_4(k,t)$ is near to unity
for large $k$, as reported in \cite{ToninelliWBBB05}; however the
exponent at small $k$ is closer to $1.5$. In the notation of
(\ref{equ:chi4_scaling_largek}) we have $N_\mathrm{p}(t)\sim
[1-P(t)]\sim t^{1/2}$ at large $k$ which accounts for the exponent of
unity in that regime.  The inset to the bottom panel shows the
function $g(r,t)$, for $r=0,1,2$.  As $t$ increases the distance $r$
over which these correlations are significant also increases.  This is
consistent with our assumption of an increasing function $N_g(t)$ in
section \ref{sec:analysis}.

\begin{acknowledgments}

We have benefited from discussions on this topic with David Reichman,
Ludovic Berthier, Hans Andersen, Phill Geissler, Dan Goldman, and Eric
Weeks.  This work was supported by EPSRC grants no.\ GR/R83712/01 and
GR/S54074/01 (JPG); University of Nottingham grant no.\ FEF 3024
(JPG); NSF grant CHE-0543158 (RLJ); and by the US Department of Energy
Grant no.\ DE-FG03-87ER13793 (DC).

\end{acknowledgments}

\appendix*

\section{Ensemble dependence of $\chi_4$}

Recently, Berthier \emph{et~al.} \cite{BerthierBBCMLLP05} suggested
that the difference between $\chi_4(k,t)$ in two different ensembles
provides a good estimate of the value of $\chi_4(k,t)$ itself.  In
this Appendix we discuss the ensemble dependence of the four point
susceptibility in the models that we consider.  Not surprisingly, we
find that the difference in $\chi_4(k,t)$ between ensembles has a
significant wavelength dependence.  At the lowest temperature we
simulated for the WCA mixture, and the highest density for the
(2)-TLG, we find that the method of \cite{BerthierBBCMLLP05} does not
give accurate numerical estimates of the susceptibility, although it
does give the correct order of magnitude for $\chi_4(k,t)$ at times
for which that function is near its maximum, the estimate working
better for smaller $k$ than for larger $k$.  It would be interesting
to see whether the estimate of \cite{BerthierBBCMLLP05} becomes better
or worse with decreasing temperature, but this is beyond the scope of
this appendix.

In terms of equation (\ref{equ:def_chi4}), the ensemble dependence of
$\chi_4(k,t)$ is contained in the definition of the average required
by the angle brackets.  This average is over any stochasticity in the
dynamics, and over an initial condition that is specified by the
choice of ensemble.  The key point is that $\chi_4(k,t)$ measures
fluctuations in an observable that couples with equal weight to every
particle in the system. So the presence of a global constraint (on the
total energy of the system, for example) suppresses these
fluctuations.  By contrast, the quantity $G_4(\bm{q},\bm{k},t)$
measures fluctuations on finite length scales, as long as $\bm{k}$ and
$\bm{q}$ are both finite.  In that case, taking the thermodynamic
limit leads to a value of $G_4(\bm{q},\bm{k},t)$ that is independent
of global constraints.

The effect of global constraints on $\chi_4(k,t)$ can be measured
directly in the (2)-TLG, by considering the susceptibility in both
canonical and grand canonical ensembles.  We denote the canonical
susceptibility by $\chi_4^{(N)}(k,t)$, which is defined as in
(\ref{equ:def_chi4}), with the averages over the dynamics, and over
all initial conditions with exactly $N$ particles.  For grand
canonical systems, we promote $N$ to a fluctuating quantity
(operator), and write:
\begin{equation}
\chi_4^{(\mu)}(k,t) = \langle \hat{N} \rangle^{-1} \Big\langle
\Big|\sum_{j=1}^{\hat{N}} \big(\hat{F}_j(\bm{k},t) -
F_\mathrm{s}(k,t)\big) \Big|^2 \Big\rangle .
\end{equation} 
where the average is over the dynamics, and over initial conditions
with all particle numbers, with weights set by the chemical potential
$\mu$.

\begin{figure}[t]
\epsfig{file=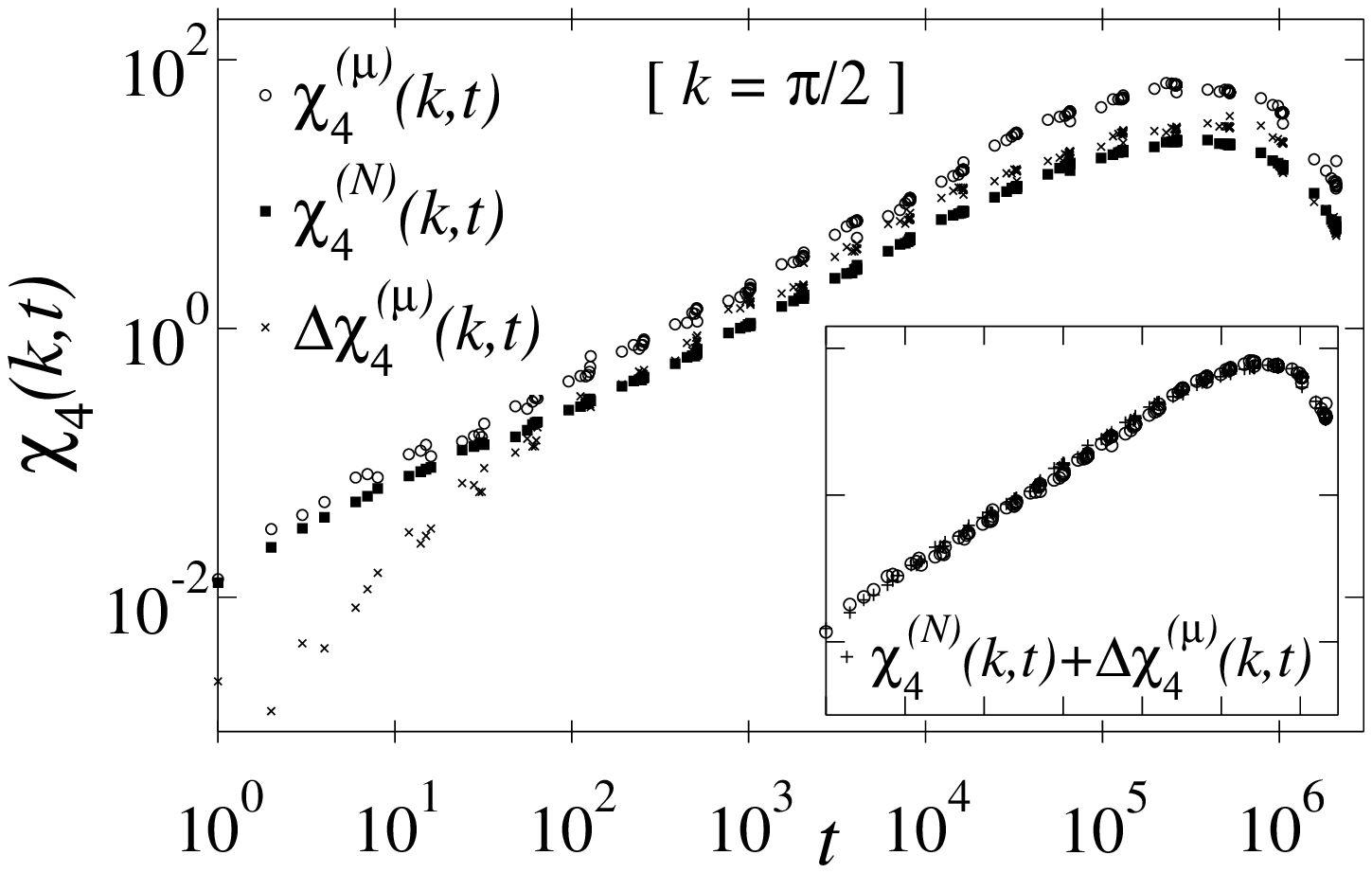,width=0.9\columnwidth}
\epsfig{file=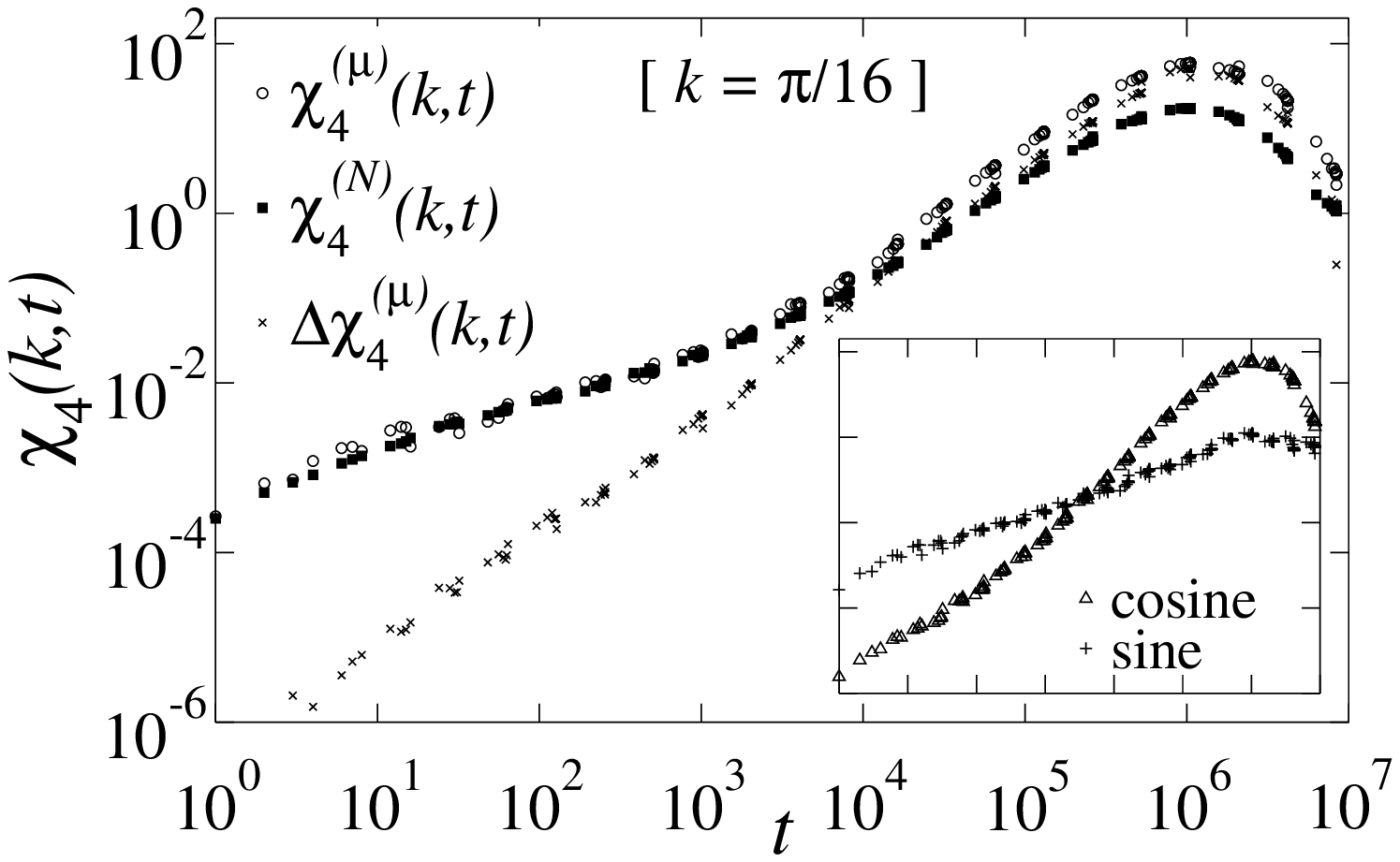,width=0.9\columnwidth}
\caption{(a) Figure showing $\chi_4(k,t)$ in the (2)-TLG at
$\rho=0.77$ and $k=\pi/2$ in both canonical and grand canonical
ensembles, together with the difference term
$\Delta\chi_4^{(\mu)}(k,t)\equiv\rho^2(1-\rho) [\partial
F_\mathrm{s}(k,t)/\partial \rho]^2$.  The contribution of
$\Delta\chi_4^{(\mu)}(k,t)$ to $\chi^{(\mu)}_4(k,t)$ is comparable to
that of $\chi_4^{(N)}(k,t)$.  (Inset, a) Illustration that the grand
canonical susceptibility satisfies the relation
(\ref{equ:chi_mu}). The scales are those of the main panel.  (b)
Similar data in the (2)-TLG at $k=\pi/16$. For smaller wave vectors,
such as these, the difference term dominates the peak of the grand
canonical susceptibility. (Inset, b) Two contributions to the
grand canonical susceptibility: see (\ref{equ:cos_sin}).  At early
times, the susceptibility is dominated by the ensemble independent
sine part, while the peak is dominated by the cosine part. The scales
are again those of the main panel.}
\label{fig:tlg_ens}
\end{figure}

\begin{figure}[t]
\epsfig{file=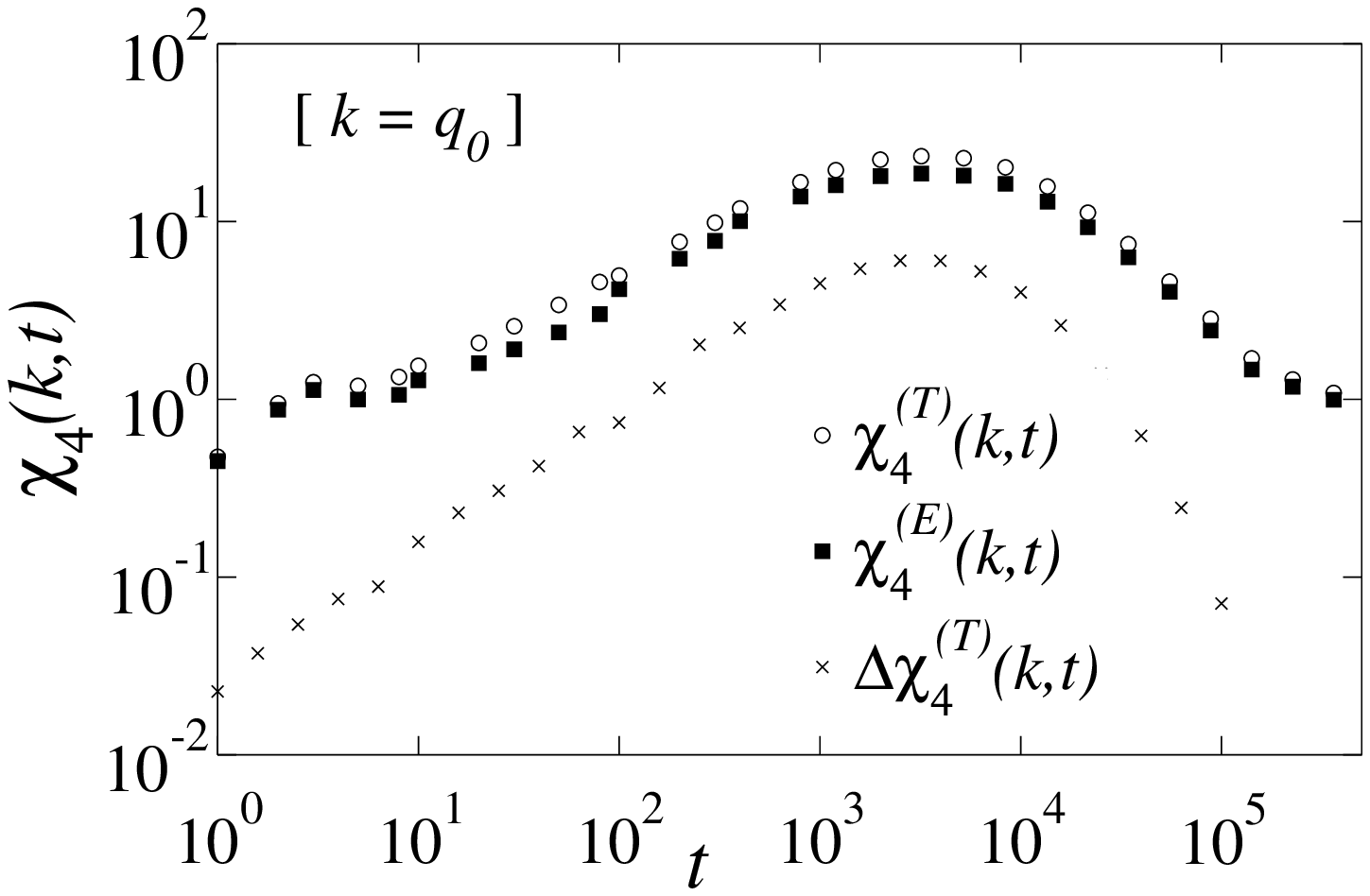,width=0.9\columnwidth}
\epsfig{file=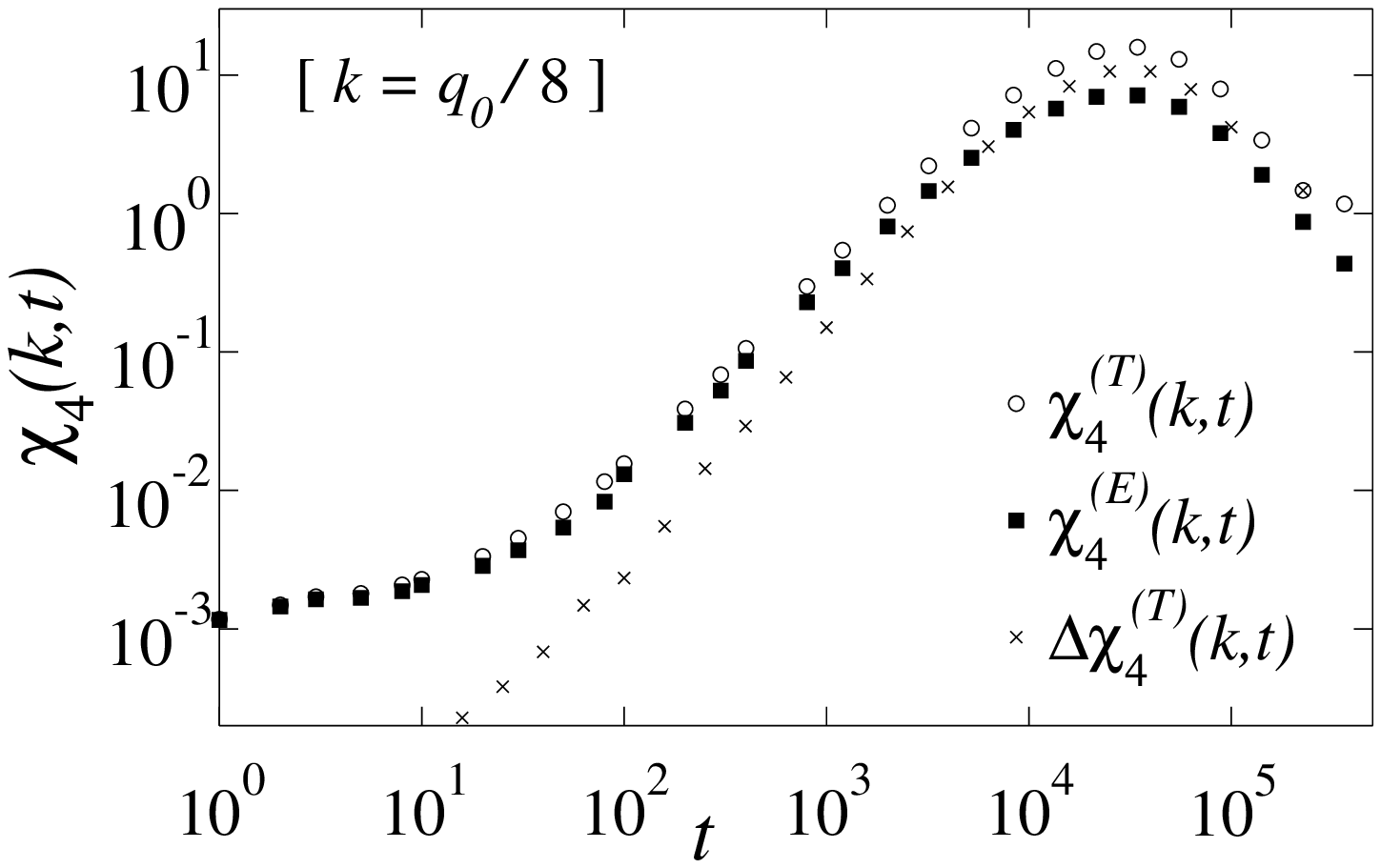,width=0.9\columnwidth}
\caption{(a) Figure showing four point susceptibilities in the WCA
mixture at $T=0.36$ and $k=q_0$, and the difference term
$\Delta\chi_4^{(T)}(k,t)$. The difference term is significantly
smaller than the canonical susceptibility.  (b) Similar data at
$k=q_0/8$. In this case the difference term exceeds the microcanonical
susceptibility. The data is consistent with the exact
relation~(\ref{equ:chi4_DeltaE}). }
\label{fig:wca_ens}
\end{figure}

These susceptibilities are related to each other by
\cite{LebowitzPV67}
\begin{eqnarray}
\label{equ:chi_mu}
\chi_4^{(\mu)}(k,t) &=& \chi_4^{(N)}(k,t) + \Delta\chi_4^{(\mu)}(k,t) 
,\\
\Delta\chi_4^{(\mu)}(k,t) &\equiv&  \rho V
\langle (\delta \rho)^2 \rangle
\left[\frac{\partial F_\mathrm{s}(k,t)}{\partial \rho}\right]^2 
,
\end{eqnarray}
where $\rho$ is the density and $V$ the volume of the system; the
average is grand canonical. The density fluctuations of the (2)-TLG
are those of a lattice gas, so that $V\langle(\delta \rho)^2
\rangle=\rho(1-\rho)$.

The relative sizes of the two terms on the right hand side of
(\ref{equ:chi_mu}) depend on the strength of the coupling of the
dynamics to density fluctuations in the system. Berthier \emph{et al.}
argued \cite{BerthierBBCMLLP05} that the second term tends to dominate
the grand canonical susceptibility in athermal glassy systems.  Since
the thermodynamic properties of the (2)-TLG system are trivial,
$\chi_4(k,t)$ can be measured in both ensembles, and the prediction
for the difference can also be evaluated. Results are shown in
figure~\ref{fig:tlg_ens}, at a high density $\rho=0.77$. The
difference between ensembles is consistent with the prediction of
(\ref{equ:chi_mu}).  We find that the two contributions to
$\chi_4^{(\mu)}(k,t)$ are of similar size for large wave vectors,
while the second term dominates at small wave vectors. In both cases,
the form of the second term is similar to that of
$\chi_4^{(\mu)}(k,t)$ for times near the peak of the susceptibility.

Our definition of $\chi_4(k,t)$ includes both the sine and cosine
parts of the operator $\hat{F}_j(k,t) =
\cos[\bm{k}\cdot\Delta\hat{\bm{r}}_j(t)] + i
\sin[\bm{k}\cdot\Delta\hat{\bm{r}}_j(t)]$, leading to two
contributions to $\chi_4(k,t)$. That is,
\begin{eqnarray}
\chi_4(k,t) &=& \chi_4^\mathrm{cos}(k,t) + \chi_4^\mathrm{sin}(k,t),
\label{equ:cos_sin}
\end{eqnarray}
with 
\begin{eqnarray}
\chi_4^\mathrm{cos}(k,t)&=& \langle\hat{N}\rangle^{-1}
\Big\langle\Big\{\sum_{j} \cos[\bm{k}\cdot\Delta\hat{\bm{r}}_j(t)] -
F_\mathrm{s}(k,t) \Big\}^2 \Big\rangle \nonumber\\ \\
\chi_4^\mathrm{sin}(k,t)&=& \langle\hat{N}\rangle^{-1} \Big\langle
\Big\{\sum_{j} \sin[\bm{k}\cdot\Delta\hat{\bm{r}}_j(t)] \Big\}^2
\Big\rangle,
\end{eqnarray}
The fluctuations in the sine part are ensemble independent, since the
average of this part is zero. Hence, the difference term
$\Delta\chi_4^{(\mu)}(k,t)$ is bounded above by the cosine part of
$\chi_4^{(\mu)}(k,t)$; it follows that the difference term is
qualitatively different from the total susceptibility when the total
susceptibility is dominated by the sine part. This is the case at
small wave vectors and small times, as shown in the inset of the lower
panel of figure~\ref{fig:tlg_ens}.

We can make a similar analysis of the WCA mixture. For each
temperature, the data of section~\ref{sec:wca} was obtained using
microcanonical simulations at eight different values of the energy,
sampled from the appropriate canonical ensemble.  This allows
estimates of both microcanonical and canonical susceptibilities, which
we denote by $\chi_4^{(E)}(k,t)$ and $\chi_4^{(T)}(k,t)$ respectively.
The microcanonical susceptibility for a given energy is given by
(\ref{equ:def_chi4}), with microcanonical averages throughout.  In
particular, we must use
\begin{equation}
\delta \hat{F}_j(k,t) = \hat{F}_j(k,t) - \langle \hat{F}_j(k,t)
\rangle_\mathrm{microcanonical} ,
\end{equation}
where the average involves data at a single energy.  Having used this
prescription to calculate $\chi_4^{(E)}(k,t)$ for each energy, we then
average $\chi_4^{(E)}(k,t)$ over the eight representative energies at
each temperature. Since the fluctuations in $\chi_4^{(E)}(k,t)$ itself
are small, the resulting average is an unbiased estimate of the
microcanonical susceptibility at the (canonical) average energy
associated with that temperature. For the canonical susceptibility, we
use (\ref{equ:def_chi4}), with canonical averages throughout,
including
\begin{equation}
\delta \hat{F}_j(k,t) = \hat{F}_j(k,t) - \langle \hat{F}_j(k,t)
\rangle_\mathrm{canonical} .
\end{equation}
We estimate canonical averages by averaging over the data from the
eight energies that are representative of the relevant temperature.

The canonical and microcanonical susceptibilities satisfy
\cite{LebowitzPV67}
\begin{eqnarray}
\chi_4^{(T)}(k,t) &=& \chi_4^{(E)}(k,t) + \Delta\chi_4^{(T)}(k,t) \\
\Delta\chi_4^{(T)}(k,t) &=& (k_B T^2/c_V) (\partial
F_\mathrm{s}(k,t)/\partial T)^2
\label{equ:chi4_DeltaE}
\end{eqnarray}
where $c_V$ is the specific heat per particle at constant volume.  The
behavior of $c_V$ that we have computed for the WCA mixture is
unremarkable.  It rises slowly with lowering temperature, indicative
of mean-square potential energy and kinetic energy fluctuations that
are virtually independent of temperature over the lowest temperatures
we have considered.

We show results for both susceptibilities in figure~\ref{fig:wca_ens},
together with an estimate of $\Delta\chi_4^{(T)}(k,t)$. [In evaluating
$\Delta\chi_4^{(T)}(k,t)$, the derivative and specific heat are
evaluated by a finite difference analysis of the temperature
dependence of the the mean energy and $F_\mathrm{s}(k,t)$].  The data
is at the low temperature $T=0.36$.  At large $k$, $\chi_4^{(T)}(k,t)$
is not saturated by the difference term
$\Delta\chi_4^{(T)}(k,t)$. Instead the microcanonical contribution is
larger. However, at smaller $k$, the difference term exceeds the
microcanonical part for times near the peak. Just as in the (2)-TLG,
the difference term becomes more significant at smaller $k$, and at
later times.  Since the size of the difference term reflects coupling
between energy fluctuations and dynamics, it would appear that motion
over large distances couples strongly to the local value of the energy
while local rearrangements do not.

Very recently, Szamel and Flenner \cite{Szamel06} measured the
contribution of global energy fluctuations to the canonical
susceptibility $\tilde{\Delta} \chi_4^{(T)}(k,t) = (k_B
T^2/c_V^\mathrm{pot})(\partial F_\mathrm{s}(k,t)/ \partial T)^2$, in a
model glass-former with Brownian dynamics (here, $c_V^\mathrm{pot}$ is
the potential energy contribution to the heat capacity, at constant
volume).  Since Brownian dynamics are non-conservative,
$\tilde{\Delta}\chi_4^{(T)}(k,t)$ does not represent a bound on the
canonical $\chi_4(k,t)$. However, the results of \cite{Szamel06} show
that while $\tilde{\Delta}\chi_4^{(T)}(k,t)$ and the unconstrained
susceptibility are not equal, they are of the same order of magnitude
for large $k$. We find a similar situation in the (2)-TLG; in the WCA
mixture then the difference term $\Delta\chi_4^{(T)}(k,t)$ is much
smaller than the canonical susceptibility at these wave vectors.


\begin{thebibliography}{99}
\bibitem{cosine-footnote} In, e.g., Ref.~\cite{ToninelliWBBB05}, only
the fluctuations of the real part of $\delta F(k,t)$ were considered.
We prefer to consider the full phase factor since both real and
imaginary parts contribute to the Fourier transform of the density;
both parts contribute to our $\chi_4(k,t)$.  Numerical contributions
of the two components of of $\chi_4(k,t)$ are similar, except when
either of $k$ or $t$ is very large or very small.
\end{thebibliography}
\end{document}